\documentstyle[12pt,aps,graphicx]{revtex} 
   \newcommand{\doublint}{\int\rule{-3.5mm}{0mm}\int} 
   \newcommand{\cent}[1] {\begin{center}#1\end{center}}
   \newcommand{\vecbm}[1]{\mbox{\boldmath#1}}
   \newcommand{\vecb}[1]{\mbox{\bf#1}}
   \newcommand{\lora} {$\longrightarrow$}
   \newcommand{\mlora} {\longrightarrow}
\tighten 
\begin{document} 
\title{ 
Fragmentation Phase  Transition in Atomic Clusters IV\\  ---
Liquid--gas transition in finite  metal clusters\\ and in the bulk
---} 
\author{ 
D.H.E. Gross and M.E.Madjet} 
\address{ 
Hahn-Meitner-Institute 
Berlin, Bereich 
Theoretische Physik, Glienickerstr.100 \\14109 Berlin, Germany\\ 
and\\ 
Freie 
Universit\"at Berlin, Fachbereich 
Physik} 
\maketitle 
\cent{\today} 
\begin{abstract} 
Within the micro-canonical ensemble it is well possible to identify
phase-transitions in small systems. The consequences for the understanding
of phase transitions in general are discussed by studying three realistic
examples. 

We present micro-canonical calculations of the fragmentation phase
transition  in Na-, K-, and Fe- clusters of $N=$ 200 to 3000 atoms at a
constant pressure of 1  atm.  The transition is clearly of first order
with a back-bending  micro-canonical caloric curve $T_P(E,V(E,P))=\partial
S(E,V(E,P))/\partial E|_P$. From the
Maxwell construction of  $\beta_P(E/N,P)=1/T_P$ one can {\em simultaneously}
determine the transition temperature $T_{tr}$, the  specific latent heat
$q_{lat}$, and the specific entropy-loss $\Delta s_{surf}$ linked  to the
creation of intra-phase surfaces. $T_{tr}\Delta s_{surf}*N/(4\pi
r_{ws}^2N_{eff}^{2/3})=\gamma$ gives the surface tension $\gamma$. Here $4\pi
r_{ws}^2N_{eff}^{2/3}=\sum N_i*4\pi r_{ws}^2m_i^{2/3}$ is the combined surface
area of all fragments with a mass $m_i\ge 2$ and multiplicity $N_i$.  All these
characteristic parameters are for $\sim$1000 atoms similar to their
experimentally known bulk values.  This finding shows clearly that within
micro-canonical thermodynamics phase transitions can unambiguously be
determined {\em without invoking the thermodynamic limit}. However, one has
carefully to distinguish observables which are defined for each
phase-space {\em point}, like the values of the conserved quantities, from
thermodynamic quantities like temperature, pressure, chemical
potential, and also the concept of pure phases, which refer to the {\em volume}
of the energy shell of the N-body phase-space and thus do not refer to a single
phase-space point.

At the same time we present here the first successful {\em microscopic
calculation of the surface tension} in liquid sodium, potassium, and iron at a
constant pressure of 1 atm..
\end{abstract} 
\section{Introduction} 
Thermodynamics is commonly believed to address to large many-body systems close
to the thermodynamic limit. It is further common believe that the three popular
ensembles, e.g. for simplicity here at given pressure $P$, the micro-canonical
\{$E,N,P$\}, the canonical \{$T,N,P$\}, and the grand-canonical \{$T,\mu,P$\}
agree in that limit and describe the same physical behaviour. But this is not
so at first order phase transitions as e.g. the liquid--gas transition.  Fixing
the temperature to the boiling temperature but leaving the energy undetermined
the canonical ensemble is a mixture of pure liquid and pure gas configurations
with equal probabilities. Drawing different members of the canonical ensemble,
the specific heat fluctuates by the amount of the specific latent heat. In the
canonical representation one would not be able to see a pot of boiling water
with a surface dividing the liquid from the vapor.  This is only possible when
the system is confined to a given specific energy $\varepsilon=E/N$ i.e. in the
micro-canonical ensemble. The energy controls what portion of the water is
liquid and what is steam.  Every-day macroscopic experience is represented by
the micro-canonical ensemble because the energy supply is seldomly
unrestricted.

Micro-canonical thermodynamics explores the topology of the N-body
phase space and determines how its volume 
\begin{equation}
W(E,N,V)=e^S=\int_{E-\delta E}^{E} dE'\doublint{
\frac{d^{3N-6}x\;d^{3N-6}p}{(2\pi\hbar)^{3N-6}}
\;\delta(H(x_i,p_i)-E')}
\end{equation} 
depends on the fundamental globally conserved quantities namely the
total energy $E=N*\varepsilon$, the angular momentum $\vecb{L}$, the
mass (number of atoms $N$), charge $Z$, the linear momentum
$\vecb{p}$, and last not least the available spatial volume $V$ of the
system.  This definition is the basic starting point of any
thermodynamics since Boltzmann\cite{boltzmann}. It is an entirely {\em
mechanistic} definition. If we do not know more about a complicated
interacting N-body system but the values of its globally conserved
macroscopic parameters the probability to find it in a special phase
space point (N-body quantum state) is uniform in the accessible phase
space. It is of course a completely separated and difficult question,
outside of thermodynamics, if and how a complicated interacting
many-body system may explore its entire accessible phase space.  This
question does not concern us here. The present work is an attempt to
develop the thermodynamics of realistic systems entirely from their
mechanics without invoking any additional assumption like the use of
the thermodynamic limit.

Before we proceed, we have to emphasise the concept of the statistical
{\em ensemble}. Each phase space cell of size $(2\pi\hbar)^{3N-6}/\delta
E$ corresponds to an individual configuration (event) of our
system. (The factor $\delta E$ is the arbitrarily chosen small width
of the energy shell. In the following all energy variables, also the
temperature, are in units of $\delta E$, here in $1$eV.)  Clearly, the
volume $e^S$ of the phase space is the sum (ensemble) of all possible
phase space cells compatible with the values of energy etc..  While
the conserved, excessive quantities, energy, momentum, number of
particles, and charge can be determined for each individual
configuration of the system, i.e.  at each phase-space {\em point} or
each event, this is not possible for the phase space {\em volume}
$e^S$, i.e. the entropy $S(E,V,N)$ and all its increments like the
temperature $T=(\partial S(E,V,N)/\partial E)^{-1}$
\footnote{)This is the thermodynamical definition of temperature. It is in
general different from its mechanical one as $2/3$ of the mean kinetic
energy per particle. This is also true for the pressure.}), the
pressure $P(E,V,N)=T\partial S(E,V,N)/\partial V$, and the chemical
potential $\mu =-T\partial S(E,V,N) /\partial N$. Their determination
demands a measure of the phase-space volume $e^S$ or its resp.
variation.  They are {\em ensemble averages}. Only in the
thermodynamic limit, for systems with infinitely many particles $N$
may e.g. the temperature be determined in a single configuration by
letting the energy flow into a small thermometer.  For a finite
system, e.g.  a finite atomic cluster, the temperature, its entropy,
its pressure can only be determined as ensemble averages over a large
number of individual events. E.g.  in a fusion of two nuclei one
obtains the excitation energy in each event from the ground-state
Q-values plus the incoming kinetic energy whereas the temperature of
the fused compound nucleus is determined by measuring the kinetic
energy spectrum of decay products which is an average over many
decays.  It is immediately clear that the size of $S$ is a measure of
the {\em fluctuations} of the system.

In what follows we discuss the most dramatic phenomena in thermodynamics: the
occurrence of phase transitions. We will try to interpret them
micro-canonically as peculiarities of the N-body phase space. We will avoid the
concept of the thermodynamic limit as we believe that this is not really
essential for the understanding of phase transitions.  We will see that details
about the transitions become more transparent in finite systems.  Then however,
one needs a modified definition for the concept of phase transitions.

In the first paper of this series on micro-canonical thermodynamics
and fragmentation of atomic clusters (papers I --- III,
\cite{gross154,gross151,gross152}) we introduced a new criterion of phase
transitions, which avoids any reference to the thermodynamic limit and
can also be used for finite systems: The anomaly of the
micro-canonical caloric equation of state $T(E/N)$ \footnote{)It is
convenient to consider specific quantities as the energy
$\varepsilon=E/N$ or the specific entropy $s=S/N$ per atom. In many
cases the micro-canonical caloric curve $T(\varepsilon)$ depends only
weakly on the particle number $N$ and the extrapolation to the
thermodynamic limit is then easy.}) \cite{gross154} where $\partial
T/\partial E \le 0$ i.e.  where the familiar monotonic rise of the
temperature with energy is interrupted. Very early the anomaly of the
caloric curve $T(E/N)$ was interpreted as signal for a phase
transition in small systems in the statistical theory of
multi-fragmentation of hot nuclei by Gross and collaborators see
refs. \cite{gross72,gross75} and the review article
\cite{gross95}. Challa and Hetherington came to the same conclusion at
about the same time in their paper on the Gaussian ensemble
\cite{challa88}. Within the Gaussian ensemble one can transform
smoothly from the micro-canonical to the canonical ensemble.  They
concluded that there is a hierarchy of ensembles : micro-canonical
$\longrightarrow$ canonical $\longrightarrow$ grand-canonical in which
the content of information about the system diminishes from left to
right. They claim it is impossible to recover the fundamental
micro-canonical ensemble from the canonical. I.e. if the canonical
partition sum is only known with limited accuracy, may be due to some
approximation, then the back-transform is dangerously inaccurate.
This is a fortiori the case at the most interesting points: At phase
transitions of first order, see below. Of course, this is because of
the high sensitivity of the Laplace--back transform from the canonical
to the micro-canonical partition sum. (E.g. as discussed below, the
Laplace transform E\lora T suppresses configurations with liquid--gas
phase separation by the exponentially small factor
exp$(-N^{2/3}\sigma_{surf}/T_{tr})$. The Laplace-back transform must
consequently enhance these configurations again by the inverse of this
factor.)  For the case of the nuclear level-density formula by Bethe
this was demonstrated by ref.\cite{gross124}.  Bixton and Jortner
\cite{bixton89} linked the back-bending of the micro-canonical caloric
curve to strong bunching in the quantum level structure of the
many-body system i.e. a sudden opening of new phase space when the
energy rises.  Their paper offers an interesting analytical
investigation of this feature.

\begin{figure}
\includegraphics*[bb = 7 168 480 514, angle=-90, width=11cm,  
clip=true]{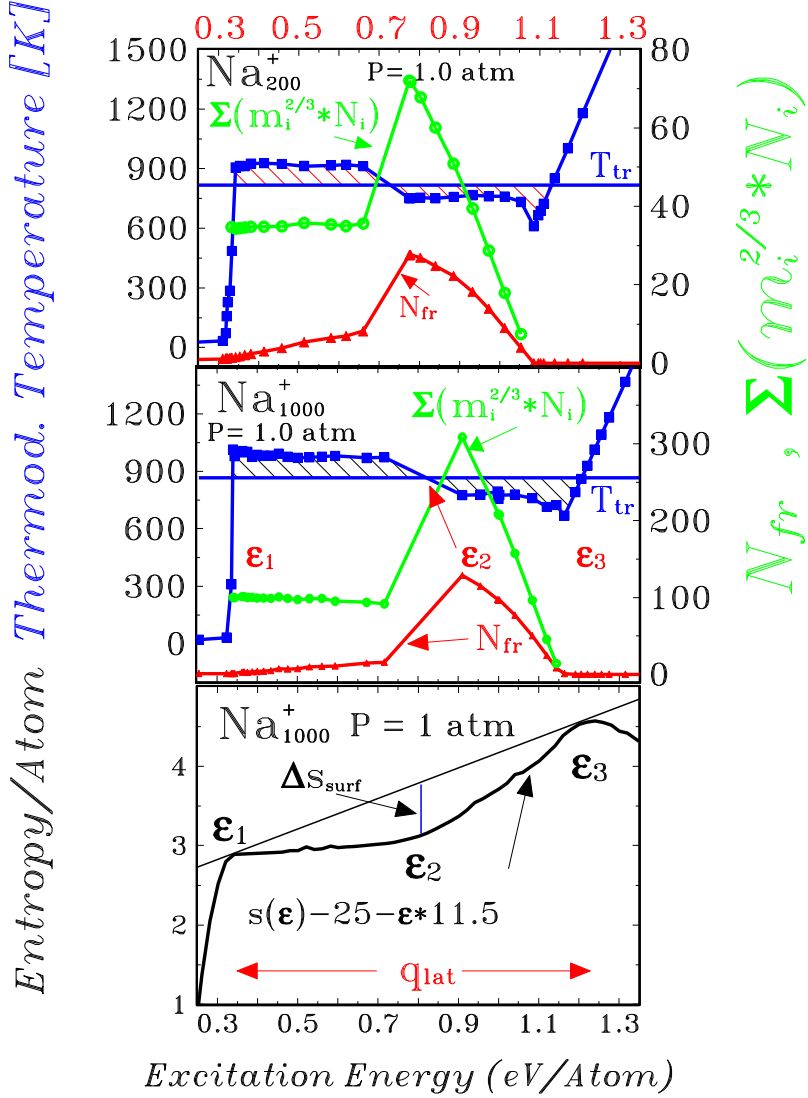}
\caption{Micro-canonical caloric curve $T_P(E/N=\varepsilon,V(E,P,N),N)$ at
constant pressure (full square points), number of fragments $N_{fr}$
with $m_i\ge 2$ atoms, and the effective number of surface atoms
$N_{eff}^{2/3}=\sum m_i^{2/3} N_i=$ total surface area divided by
$4\pi r_{ws}^2$. The two shaded areas correspond to the two equal
areas between $\beta(\varepsilon)=1/T(\varepsilon)$ and the
Maxwell-line $\beta= 1/T_{tr}$. In the lower panel
$s(\varepsilon)=\int_0^{\varepsilon}\beta(\varepsilon')d\varepsilon'$
is shown.  In order to make the intruder between $\varepsilon_1$ and
$\varepsilon_3$ visible, we subtracted the linear function
$25+11.5\varepsilon$. The double-tangent to $s(\varepsilon)$ is the
concave hull to $s(\varepsilon)$ between $\varepsilon_1$ and
$\varepsilon_3$. Its derivative is the Maxwell-line in the upper two
panels.
\label{fig1}}
\end{figure}

A phase transition of first order is characterised by a sine-like
oscillation, a ``back-bending'' of $T(\varepsilon=E/N)$
c.f. fig.\ref{fig1}. As shown below, the Maxwell-line which divides
the oscillation of $\partial S/\partial E=\beta(\varepsilon)=1/T$ into
two opposite areas of equal size gives the inverse of the transition
temperature $T_{tr}$, its length the specific latent heat $q_{lat}$,
and the shaded area under each of the oscillations is the loss of
specific entropy $\Delta s_{surf}$. The latter is connected to the
creation of macroscopic intra-phase surfaces, which divide mixed
configurations into separated pieces of different phases, e.g.  liquid
droplets in the gas or gas bubbles in the liquid.  Even nested
situations are found in some other cases like liquid droplets inside
of crystallised pieces which themselves are swimming in the liquid in
the case of the solid -- liquid transition, see e.g.  the experiments
reported in \cite{grimsditch96}.  I.e. at phase transitions of first
order inhomogeneous ``macroscopic or collective'' density fluctuations
are common, boiling water is certainly the best known
example. Phase-dividing surfaces of macroscopic size exist where many
atoms collectively constitute a boundary between two phases which
cause the reduction of entropy by $\Delta s_{surf}$.  As the entropy
is the integral of $\beta(\varepsilon)$:
\begin{equation}
s(\varepsilon)=\int_0^{\varepsilon}{\beta{(\varepsilon')}d\varepsilon'}
\end{equation}
it is a concave function of $\varepsilon$
($\partial^2s/\partial\varepsilon^2 =\partial\beta/\partial\varepsilon
< 0$) as long as $T(\varepsilon)=\beta^{-1}$ shows the usual monotonic
rise with energy. In the pathological back-bending region of
$\beta(\varepsilon)$ the entropy $s(\varepsilon)$ has a convex
intruder of depth $\Delta s_{surf}$ \cite{gross150}(c.f.
fig.\ref{fig1}c). This has the consequence that the specific heat
$(\partial T/\partial E)^{-1}$ becomes {\em negative}.  This
astonishing result was discussed early by Thirring \cite{thirring70}.
At the beginning ($\ge\varepsilon_1$) of the intruder the specific
entropy $s(\varepsilon)$ is reduced compared to its concave hull,
which is the tangent to $s(\varepsilon)$ in the points $\varepsilon_1$
and $\varepsilon_3$. The derivative of the hull to
$\beta(\varepsilon)$ follows the Maxwell-line in the interval
$\varepsilon_1\le\varepsilon\le\varepsilon_3$.  In the middle,
($\varepsilon_2$), when the separation of the phases is fully
established this reduction is maximal $=\Delta s_{surf}$ and at the
end of the transition ($\varepsilon_3$) when the intra-phase surface
disappears $\Delta s_{surf}$ is gained back.  Consequently, the two
equal areas in $\beta(\varepsilon)$ are the initial loss of surface
entropy $\Delta s_{surf}$ and the later regain of it.  Due to van
Hove's theorem \cite{vanhove49} this convex intruder of
$s(\varepsilon)$ must disappear for ``thermodynamically stable
systems'' as we consider in this paper in the thermodynamic limit
which it will do if $\Delta s_{surf} \sim N^{-1/3}$. This is why a
transition of first order may easier be identified in finite systems
where the intruder can still be seen.  The intra-phase surface tension
is related to $\Delta s_{surf}$ by $\gamma_{surf}=\Delta
s_{surf}*N*T_{tr}/\mbox{surf.-area}$.

In conventional thermodynamics a phase transition of first order is
indicated by a singularity (jump) in the specific heat $c_P(T)$ in the
thermodynamic limit ($N \mlora
\infty|_{\varrho=N/V=\mbox{\footnotesize const}}$).  It is easy to see
the relation to our present criterion. The canonical partition sum is
obtained from the micro-canonical volume $W(E)$ of the accessible
N-body phase space, in units of $(2\pi \hbar)^{3N-6}/\delta E$, or the
number of quantum mechanical N-body states, $W(E)=
e^{Ns(\varepsilon=E/N)}$ by the Laplace transform from the extensive
energy to the intensive temperature $T=1/\beta$:
\begin{equation} 
Z(\beta)=\int_0^{\infty}{W(E)e^{-\beta E}}\;dE .
\label{laplace} 
\end{equation} 
In the thermodynamic limit ($N \mlora
\infty|_{\varrho=N/V=\mbox{\footnotesize const}}$) the  partition
function $Z(\beta)$ and the bulk specific heat $c(T)$ are obtained via the
Laplace transform (\ref{laplace}) in saddle-point approximation (see e.g.
\cite{bohrmott} App.2B) :
\begin{eqnarray} 
Z(\beta)&=&\int_0^{\infty}{W(N\varepsilon)\;e^{-\beta N
\varepsilon}\;Nd\varepsilon} 
\label{laplacest}\\  
&\sim&
T\;e^{N[s(\bar{\varepsilon})-\beta\bar{\varepsilon}]}\;\sqrt{2\pi 
N/c(\bar{\varepsilon})}\label{laplace1}\\ 
\frac{\partial}{\partial\varepsilon}s(\varepsilon)|_{\bar{\varepsilon}}&=
&\beta. \label{station} \\
c(\bar{\varepsilon})&=&-\frac{\partial ^2s}{\partial\varepsilon ^2}  
\left.\frac{1}{\left(\frac{\partial s}{\partial \varepsilon}\right)^2}
\right|_{\bar{\varepsilon}} .
\end{eqnarray} 

The exponent in eq.(\ref{laplace1}) is the free energy $-Nf(T)/T$.
Eq.\ref{station} establishes a relation between $T$ and
$\varepsilon$, the caloric equation of state. Whereas
$T(\bar{\varepsilon})$ is single valued everywhere, this is not
true for its inverse $\bar{\varepsilon}(T)$ in the back-bending
region.  There are in general 3 solutions for $T=T_{tr}$ of
eq.(\ref{station}) here c.f. fig.\ref{fig1}: $\bar{\varepsilon}=$
$\varepsilon_1$, $\varepsilon_2$, and $\varepsilon_3$. At the
second solution $\varepsilon_2$ of equation (\ref{station})
$s(\varepsilon)$ has a positive curvature and the saddle point is
a minimum in the direction of integration not a
maximum. Consequently, the Laplace transform (\ref{laplacest})
towards the canonical partition function $Z(T)$ jumps over this
region which becomes exponentially suppressed by a factor
$e^{-\Delta f_{surf}/T_{tr}}$ where $\Delta f_{surf}\propto
N^{2/3}$ is the positive surface free energy. In the thermodynamic
limit ($N\mlora
\infty$) the integration follows the concave hull of
$s(\varepsilon)$ here.  Then $c(T)$ gets the singularity
$=q_{lat}\delta(T-T_{tr})$ but otherwise remains finite
\cite{hueller94,promberger96}, which is just the conventional signal
of a phase transition of first order, c.f. Ruelle
\cite{ruelle69}. This jump over the energies
($\varepsilon_1\le\varepsilon\le\varepsilon_3$) of configurations with
phase-separation with the consequence of their exponential suppression
is the mathematical reason for the loss of information in the
canonical ensemble. It is just the crucial information about the phase
transition.

Our approach to phase-transitions of first order is complementary to
the conventional approaches where the separation of the system into
two homogeneous phases by a --- in general --- geometrical interface
is investigated, e.g.\cite{panagiotopoulos87}. The problems due to the
large fluctuations of this interface are numerous and severe, see
e.g. the discussion on fluid interfaces by Evans \cite{evans92}. These
fluctuation are of course crucial for the interfacial entropy and
consequently for the surface tension also.  The two main differences
of our approach are that
\begin{itemize}
\item we do not start with the {\em geometry} (planar or spherical) of the
interface but focus our attention to the {\em entropy} of the phase separation.
This turns out to be much simpler than the geometric approach. Moreover,
it is the entropy of the interface that decides the transition not its 
geometrical interpretation.
\item The micro-canonical ensemble allows for large scale spatial
inhomogeneities, whereas the canonical ensemble {\em suppresses}
spatial large scale inhomogeneities like phase-separations
exponentially
\{in the case of a phase transition of first order $\propto
exp(-\sigma N^{2/3}/T_{tr})$, where $\sigma$ is the surface tension
parameter ($\sigma =4\pi r_{ws}^2\gamma$) in the liquid drop
parametrisation of the ground-state binding-energy of the clusters here of
course at the boiling point,
$r_{ws}$ is the radius of the Wigner-Seitz cell
\cite{gross153}\}. This is an example of the information loss in the
canonical ensemble mentioned by Challa and
Hetherington\cite{challa88}.
\end{itemize}

This characterisation of phase transitions is purely
thermodynamically. We have not yet defined what a phase is.  Much
effort is spent in Ruelle's book to define pure phases as those
configurations for which in the thermodynamic limit observables
survive increasing coarse-graining, for which space-averaged
quantities do not fluctuate, c.f.  chapter 6.5 in Ruelle's book
\cite{ruelle69}. Of course, this definition works in the thermodynamic
limit only. It does not address to finite systems. For a finite system
it is not possible to decide if a specific configuration corresponds to
a pure phase or not. The situation is analogous to the definition of
the temperature, see above. Again to be a pure phase is a feature of
the whole ensemble not of a single phase-space point (configuration).
We use a {\em statistical} definition of a pure phase: A configuration
belongs to the ensemble of pure phases --- including its fluctuations
--- at concave points of $S(E,N,V(E,P))$ with $\partial^2 S/\partial
E^2 < 0$. At the two crossing points of the micro-canonical caloric
curve $T_P(E/N)=T_{tr}$, $\varepsilon_1,\varepsilon_3$ we say the
system is in the pure ``liquid'' phase or in the pure ``gas'' phase
respectively. This is at this moment nothing more than a working
hypothesis.  However, we will see that for the systems we have
investigated (2-dim Potts models, fragmentation of sodium, potassium,
and iron clusters with $N=200-3000$ atoms) these definitions give very
realistic values for {\em all} parameters of the transitions quite
similar to the ones in the bulk. Of course, the parameters depend
weakly on the number of particles considered.  However, there is no
reason to emphasise any {\em qualitative} difference between the bulk
transition and the ``phase transition'' with or without quotation
marks in finite systems as defined above.

\begin{figure}
\begin{center}
\includegraphics*[bb = 23 50 446 612, angle=-90, width=10cm,
clip=true]{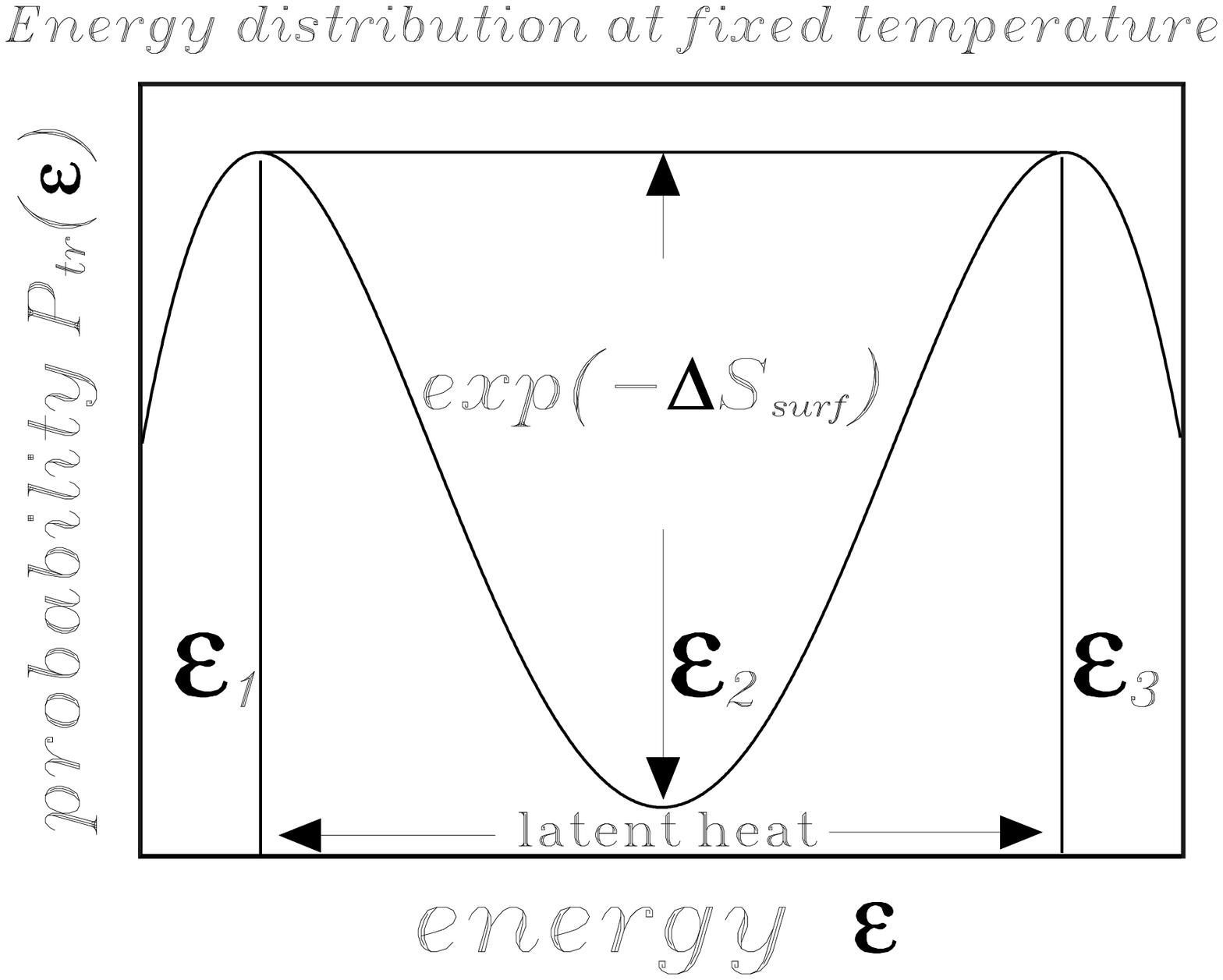}
\end{center}                                         
\caption{Probability $P_{tr}(\varepsilon)$ to find the system in the canonical 
ensemble at a phase transition of first order ($T=T_{tr}$) at the
energy/particle $\varepsilon$. At $\varepsilon =\varepsilon_1$ it is ``
liquid'' and at $\varepsilon =\varepsilon_3$ it is in the pure ``gas'' phase
\label{PvE}}
\end{figure}

It is evident, at a phase transition of first order ($T=T_{tr}$) the
members of a canonical ensemble split into two distinct groups with
different energy/particle ($\varepsilon_1$ and $\varepsilon_3$) but
with equal probability $e^{-f/T}$ , a ``liquid'' and a ``vapour''
phase because both have the same free energy. At $T_{tr}$ the energy
{\em per particle} $\varepsilon$ fluctuates by the specific latent
heat whatever the coarse-graining might be.  This is illustrated in
fig.\ref{PvE}. Consequently, {\em at a first order phase transition
the micro-canonical ensemble where energy does not fluctuate is
different from the canonical one even in the thermodynamic limit at
$T=T_{tr}$.}.

 If the shaded area $\Delta s_{surf}$ under the oscillation of
 $\beta(\varepsilon)$ (see fig.\ref{fig1}) disappears and if the latent
 heat $q_{lat}=0$, we have a continuous phase transition. In the thermodynamic
limit the Laplace transform eq. (\ref{laplacest}) then has only one stationary
point at $T_{tr}$ in this case, which is a saddle point of the caloric curve
$T(\varepsilon)$ and the specific heat $c(T)=\partial\varepsilon/\partial T$ is
continuous and has a pole at $T=T_{tr}$, c.f.  fig.1 of paper I
\cite{gross154}.

In ref.\cite{gross150} we showed for a two-dimensional Potts-model
that all the above mentioned micro-canonical parameters of a phase
transition of first order {\em are within a few percent close to their
bulk values for relatively small systems like a couple of hundreds
spins}. This is not only true in model systems but also in realistic
continuous systems like liquid metals.  In this paper (IV. of our
series) we will follow the micro-canonical fragmentation transition in
clusters of sodium, potassium, and iron with increasing number of
atoms towards the well known liquid-gas transition of the bulk. In
contrast to the fragmentation transition in isolated clusters which
have to be treated at an approximately constant volume defined by the
short range of dissipation \cite{gross153}, the liquid-gas transition
must be studied at constant pressure, here at 1 atmosphere. The rapid
convergence of the three characteristic parameters, $T_{tr}$,
$q_{lat}$, and $\sigma_{surf}$ towards their bulk values (c.f. tables
\ref{table1} and \ref{table2}) is a further check of the validity  of
our characterisation of a phase transition by the anomaly of the
micro-canonical caloric curve and of our numerical method ({\em MMMC})
to treat the fragmentation of small systems like nuclei or atomic
clusters.

The advantage of our new signal of a phase transition is the
following: As shown in \cite{hueller94,promberger96,gross150,gross153}
the micro-canonical caloric equation of state $T(E/N)$ gives this
signal of a phase transition already for relatively small
systems. Moreover, $T(E/N)$ becomes rather quickly independent of
the number of particles outside the back-bending region, which itself
degenerates for large $N$ towards the Maxwell-line $T=T_{tr}$.
\newpage
\section{The micro-canonical liquid--gas transition\\ at constant pressure}
The entropy of the system follows from the fundamental formula
\begin{equation} 
S(E,V,N) = k\ast lnW(E,V,N) , 
\end{equation} 
where $W(E,V,N)$ is the number of all accessible quantum states of the
N-body system in the energy interval $E, E+\delta E$ and volume
$V$. In what follows we take Boltzmann's constant $k=1$. The
thermodynamic temperature at constant volume is defined by (in the
following we suppress the dependence of the number of atoms $N$, as
this is hold fixed in all derivatives)
\begin{eqnarray} 
\frac {1}{T(E,V)}&=& \beta(E,V) = 
\frac {\partial S(E,V)}{\partial E} \label{T}\\ 
&=&<\frac{f_{tr}-2}{2 E_0'}>, 
\end{eqnarray} 
where $f_{tr}$ denotes the number of  translational-rotational degrees of
freedom and $E_0'$ is the remaining kinetic energy (c.f. eq.B43
\cite{gross153}).  The pressure is given by
\begin{eqnarray} 
P(E,V)&=&T(E,V) * \left( \frac {\partial S(E,V)}{\partial V}
\right)_{E} \label{P}\\ &=&T(E,V)<\frac{\partial w_m w_r}{\partial V}>\\
&&\mbox{$w_m w_r$ is the spatial weight given in eqs.B25, B30 of
\cite{gross153}:}\\ w_m w_r&=&\frac{1}{N_m!} {\left[ g_e \, {\left( \frac{1}{2
\pi \hbar} \right) }^{3} \, V_m \right] }^{N_m}
\prod_{j=1}^{N_f}\left\{\frac{1}{(2 \pi
\hbar)^3}\frac{4\pi}{3}(R_{sys}-R_j)^3\right\} /NCC\\
P&\approx&T<\frac{N_f+N_m}{V}>+\mbox{terms due to the change of $NCC$},
\label{PP}
\end{eqnarray} 
where $NCC-1$ is the number of unsuccessful attempts to put the
fragments $j$ into the given volume. It takes care of the varying
avoided volume. In the present calculations at a pressure of 1atm we
have $NCC\approx 1$ so that its variation does not concern us here. At
higher pressure, more close to the critical point, the pressure due to
the change of the avoided volume is more important.  $N_f+N_m$ is the
average number of fragments including monomers and $V$ is the
available volume of the system. (As already discussed above, in
contrast to here, for the fragmentation phase transition of free
finite clusters the calculations were done in papers I--III
\cite{gross154,gross150,gross152} at constant freeze-out volume not at
constant pressure.) We can calculate the thermodynamic temperature at
constant pressure by the following:
\begin{equation} 
\frac {1}{T_P(E,P)}= \beta_P(E,V(E,P))=\frac {\partial S(E,V(E,P))} 
{\partial E}\Big\vert_{P} = \frac {\partial (SP)} 
{\partial(EP)} =\frac{ \frac {\partial (SP)}{\partial (EV)}} 
{\frac {\partial (EP)}{\partial (EV)}} .
\end{equation} 
Using the Jacobian
\begin{equation} 
\frac{\partial (SP)}{\partial (EV)} =  
\left[\frac{\partial S(E,V)}{\partial E}\Big\vert_{V}  
\frac{\partial P(E,V)}{\partial V}\Big\vert_{E} -  
\frac{\partial S(E,V)}{\partial V}\Big\vert_{E}  
\frac{\partial P(E,V)}{\partial E}\Big\vert_{V} \right] , 
\end{equation} 
we get 
\begin{equation}
\beta_P(E,P) =  \left[ \frac{\partial S} 
{\partial E}\Big\vert_{V} \frac{\partial P} 
{\partial V}\Big\vert_{E} - \frac{\partial S}{\partial V} 
\Big\vert_{E} \frac{\partial P} 
{\partial E}\Big\vert_{V}\right]  
\frac{1}{\frac{\partial P}{\partial V}\Big\vert_E} ,
\end{equation} 
and 
\begin{equation} 
\beta_P(E,P)=\beta(E,V) \ast \left[ 1 - P \frac{\frac  
{\partial P}{\partial E}\Big\vert_{V}}{ 
{\frac{\partial P}{\partial V}\Big\vert_E}} \right] .
\end{equation} 
On the other hand we have from eqs.\ref{P},\ref{T} :
\begin{eqnarray} 
\frac{\partial P(E,V)}{\partial E}\Big\vert_{V} &=&  
\frac{\partial T(E,V)}{\partial E}\Big\vert_{V}  
\frac{\partial S(E,V)}{\partial V}\Big\vert_{E} + T \frac{\partial^2 S(E,V)} 
{\partial E \partial V} \\
\frac{\partial P(E,V)}{\partial V}\Big\vert_{E} &=&  
\frac{\partial T(E,V)}{\partial V}\Big\vert_{E} 
\frac{\partial S(E,V)}{\partial V}\Big\vert_{E} 
+ T \frac{\partial^2 S(E,V)}{\partial V^2}\Big\vert_{E}\\ 
\frac{\partial T(E,V)}{\partial E}\Big\vert_{V} &=&  
- T^2 \frac{\partial \beta(E,V)}{\partial E}\Big\vert_{V} 
= - T^2 \frac{\partial^2 S(E,V)}{\partial^2 E}\Big\vert_{V} .
\end{eqnarray} 
Using $w_m$ and $w_r$ given in the appendix B of the review paper
\cite{gross153} and assuming for the moment the avoided volume to keep close to
the initial cluster volume (at low densities of the system), we finally get
\begin{eqnarray} 
\beta_P(E,P)&=&\beta(E,V(E,P)) \left[ 1 - \frac{T <N_{t}> (< \frac{N_{t} (f_{tr} -2)}
{2E_{0}'} > - T <N_{t}> < \frac{(f_{tr} -2 )(f_{tr} -4)}{4E_{0}'^2}
>)} {(<N_t(N_t-1)> - T <\frac{(f_{tr} -2)N_t}{2E_{0}'}> <N_t>)}
\right] \nonumber\\ 
&&+\mbox{terms $\propto \partial NCC/\partial V$ at low
$\varepsilon$, high $P$},\label{betap}
\end{eqnarray} 
where $N_t$ is the total number of fragments including neutral
monomers.  $E_{0}'$ is the remaining energy and $f_{tr}$ is the
number of translational-rotational   degrees of freedom.
 
One notes that $\beta(E,V) = <\frac{(f_{tr} -2)}{2E_{0}'}>$.  
In the case of vaporisation (only monomers),  
we find 
\begin{equation} 
\beta_P(E,P)=\beta(E,V(E,P)) \left[ 1 + \frac{2N}{3N-8}\right] , 
\end{equation}
where $N$ is the number of atoms.  At very large number of atoms we get the
same formula as derived for an ideal gas consisting of $N$ particles
\begin{equation} 
\beta_P(E,P)=\beta(E,V(E,P))  \frac{5}{3} .
\end{equation} 
The corrections $\propto \partial NCC/\partial V$ take care of the fact that
the avoided volume is larger than the total eigenvolume of the fragments. It is
bigger at low excitation energies when the fragments are larger than at higher
excitation.  These variations, consequently, contribute to the pressure
(eq.\ref{PP}). These corrections turn out to be negligible at 1 atm.. They are,
however, more important at higher pressure when one approaches the critical
point.

The micro-canonical ensemble with given pressure $W(E,P,N)$ must be 
distinguished from the (in spirit) similar constant pressure ensemble \{H,P,N\}
introduced by Andersen \cite{andersen80,brown84,procacci94} where a 
molecular-dynamic calculation with the hypothetical Hamiltonian \cite{brown84}
\begin{equation}
H(\{r_i,p_i\},V,\dot{V})=\frac{V^{2/3}}{2}\sum_{i=1}^{N}m\vecbm{$\dot{r_i}$}*
\vecbm{$\dot{r_i}$} + \sum_{i=1}^N\sum_{j>1}^N\Phi(r_{ij}V^{1/3})+
\frac{M}{2}\dot{V}^2+P_EV
\end{equation}
is suggested. Here $V$ is the volume of the system, taken as an
additional explicit degree of freedom, $\{r_i,p_i\}$ are the
coordinates and momenta of the atoms scaled with the factor $V^{1/3}$,
$\Phi(r_{ij})$ is the intra-atomic two-body potential, and $M$ is a
hypothetical mass for the volume degree of freedom. $P_E$ is the
given pressure. The total ``energy'' $H$, atoms plus $V$-degree
of freedom, is conserved, not the total energy $E$ of the atoms alone. 

This is different to our micro-canonical approach with given
$E,V(E,P),N$ where the energy $E$ of the atoms is conserved and the
pressure is the correct thermodynamic pressure ($P(E,V)=T(E,V)\partial
S/\partial V|_E$). At each energy the volume $V(E,P)$ is chosen for
all members of the ensemble simultaneously by the condition that
$T(E,V)\partial S(E,V)/\partial V|_E$ of the whole ensemble is the
correct pressure. In this case there is a unique correlation between
the energy $E$ and the volume $V$ which does not fluctuate within the
ensemble even though the pressure is specified. At the given energy
this is still the \{E,V(E,P,N),N\} ensemble.  Moreover, the entropy
cannot be calculated directly by molecular dynamics, whereas this is
possible in our micro-canonical Metropolis Monte Carlo method.

\section{The liquid-gas transition of sodium, potassium, and iron} 
The microscopic simulation of the liquid--gas transition in metals is
especially difficult.  Due to the delocalisation of the conductance electrons
metals are not bound alone by two-body forces but experience long-range
many-body interactions. Moreover, at the liquid--gas transition the binding
changes from metallic to covalent binding. This fact is a main obstacle for the
conventional treatment by molecular dynamics \cite{allen85}.

In the macro-micro approach we do not follow each atom like in
molecular dynamics, the basic particles are the fragments. Their
ground-state binding energies are taken from experiments. The
fragments are spherical and have translational, rotational, and
intrinsic degrees of freedom.  The internal degrees of freedom of the
fragments are simulated as pieces of bulk matter. The internal density
of states is calculated from the internal entropy of the fragments. It
is taken as the specific bulk entropy $s(\varepsilon)$ at excitation
energies $\varepsilon\le \varepsilon_{max}=\varepsilon_{boil}$. The
bulk entropy can be determined from the experimentally known specific
heat of the solid/liquid bulk matter
\cite{gross141}. $\varepsilon_{boil}$ is the specific energy where the
boiling of bulk matter starts. This approximation allows to take
important enharmonicities of the internal degrees of freedom into
account e.g. near to the melting point.  Details are discussed in
\cite{gross153}. Then the metallic binding poses no difficulty for us and the
metal --- nonmetal transition is controlled in our approach by the increasing
fragmentation of the system. This leads to a decreasing mean coordination
number when the transition is approached from the liquid side while the
distance to the nearest neighbour remains about the same. Exactly this
behaviour was recently observed experimentally \cite{hensel95,ross96}. By using
the micro-canonical ensemble we do not pre-specify the intra-phase surface and
allow it to take any form. Also any fragmentation of the interface is allowed.
It is the entropy alone which determines the fluctuations of the interface.
Here we present the first successful microscopic calculation of the surface
tension in liquid sodium, potassium, and iron.

\begin{figure}
\includegraphics*[bb = 12 2 496 573, angle=-90, width=16cm,  
clip=true]{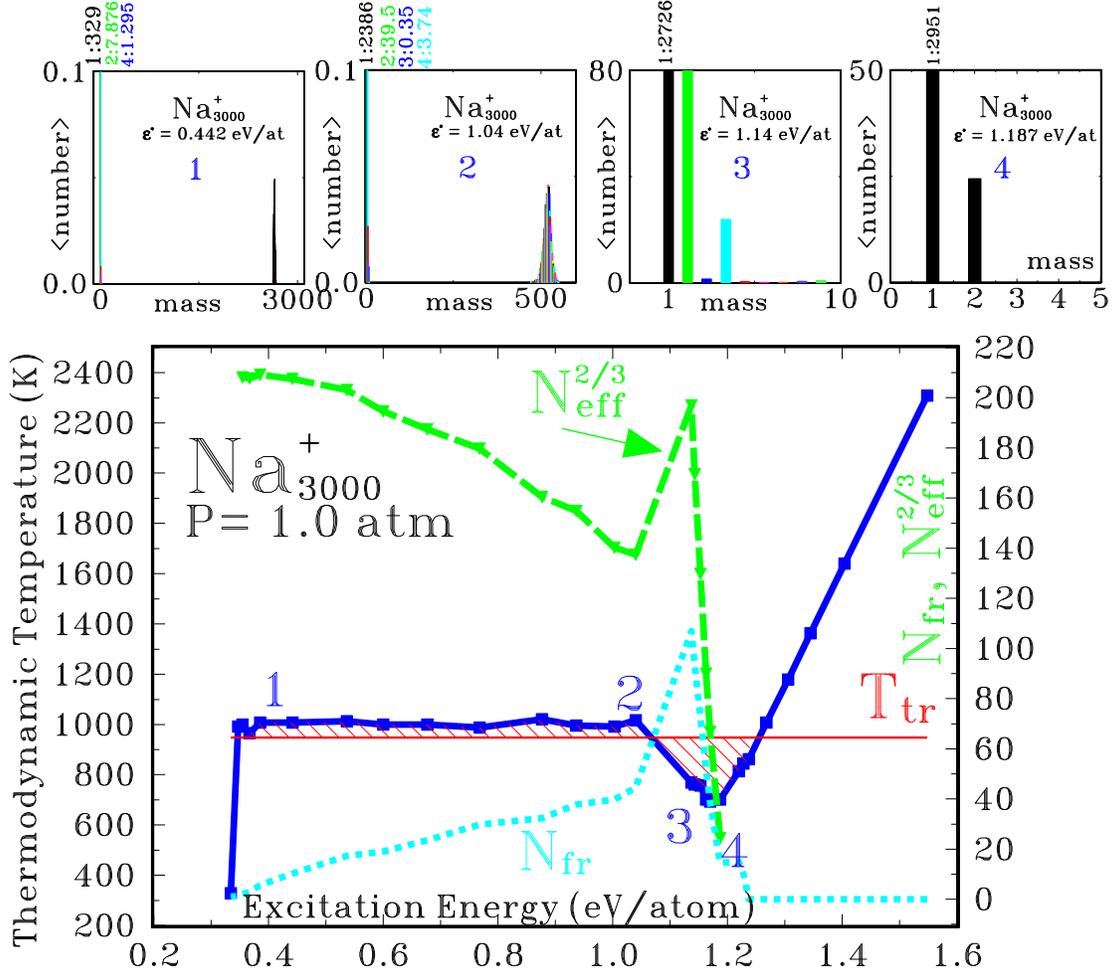}
\caption{Same as \protect\ref{fig1} but for Na$^+_{3000}$. The four small
figures at the top show the mass distribution of fragments at four different
excitation energies which are indicated in the main figure by their number.
The small vertical numbers on top of the mass-distributions give the real
number of fragments e.g.: 2:7.876 means there are 7.876 dimers on average at
$\varepsilon=0.442$eV/atom.
\label{fig1a}}
\end{figure}

\begin{figure}
\includegraphics*[bb = 38 148 485 587, angle=-90, width=14cm,  
clip=true]{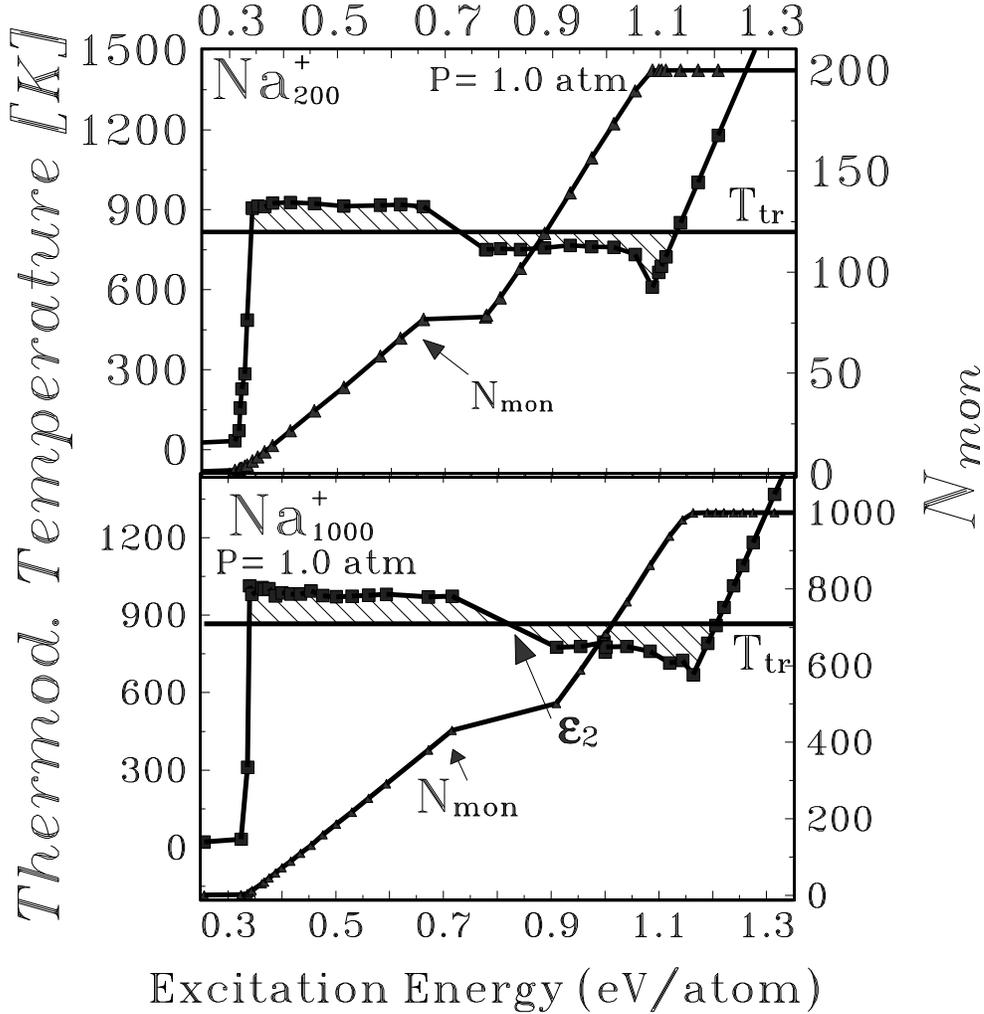}
\caption{Micro-canonical caloric curve $T_P(E/N=\varepsilon,P)$ (full square
points), number of monomers
\label{fig2}}
\end{figure}

The figure \ref{fig1} shows the micro-canonical caloric curve
$T(\varepsilon)$ for a system of $N=$ 200 and 1000 sodium atoms. The
back-bending of $T_P(\varepsilon)$ can be clearly seen. At
$\varepsilon\sim\varepsilon_2\approx 0.7-0.8$eV the thermodynamic
temperature drops suddenly due to the rapid increase of the number of
fragments. This induces a jump in the increment $\partial
s/\partial\varepsilon=1/T(\varepsilon)$ of the entropy
$s(\varepsilon)$. The number of fragments with mass $m_i\ge 2$
increases slowly up to $\varepsilon_2$ and from there on jumps up and
decays continuously down to 0. $4\pi r_{ws}^2N_{eff}^{2/3}=\sum4\pi
r_{ws}^2 m_i^{2/3}*N_i$ is the total surface area of the
fragments. For $\varepsilon \le \varepsilon_2$ the size $m_i$ of the
fragments decreases due to an increasing evaporation of monomers,
c.f. fig.\ref{fig2}, but the number of fragments increases such that
the total surface area decreases more weakly with rising
excitation. For $N_0=200$ and $1000$ it keeps even approximately
constant $\approx 4\pi r_{ws}^2N^{2/3}$. A more detailed investigation
shows that in sodium as well as also in potassium we actually may have
two interfering transitions: One from evaporation of monomers and
smaller fragments with a large residue towards multi-fragmentation
with several medium sized fragments at $\sim 0.7$eV/atom and a second
one from multi-fragmentation into a pure gas of monomers at
$\varepsilon > 1.2$eV/atom. For systems with more atoms, $N_0\ge 3000$
c.f. fig.\ref{fig1a}, the multi-fragmentation in sodium moves towards
larger excitation and melts together with the vaporisation dip at
$\sim 1.1$--$1.2$eV/atom. In the canonical ensemble all these
important details become suppressed or even hidden.
\newpage
The bulk values of $\sigma$ are calculated from the experimental
surface tension $\gamma$ by:
\begin{eqnarray}
\sigma_{bulk}&=&4\pi r_{ws}^2\gamma\\
r_{ws}^3&=&10^{30}\frac{3vM}{4\pi L}\\
\sigma_{bulk}(T)&=&6.242*10^{-5}*4\pi
(r_{ws}^2\gamma)|_{T_{melt}}\left\{1+\left[\left.\frac{1}
{\gamma}\frac{d\gamma}{dT}\right|_{T_{melt}}+
\left.\frac{2}{3v}\frac{dv}{dT}\right|_{T_{melt}}\right](T-T_{melt})\right\}
,\label{sigmabulk}
\end{eqnarray}
where $r_{ws}$ in [$\AA$] is the Wigner-Seitz radius at melting, $v$
in [$\frac{m^3}{kg}$] is the specific volume at melting, $L$ is the
number atoms per kg-mole (Loschmid's number), and $M$ is the molecular
weight.  $\gamma|_{T_{melt}}$, $d\gamma/dT|_{T_{melt}}$ are the
experimental surface tension in [$\frac{\mbox{mN}}{\mbox{m}}$] and
$v|_{T_{melt}}$, $\frac{dv}{dT}|_{T_{melt}}$ are the specific volume
in [$\frac{m^3}{\mbox{kg}}$] and its temperature derivative in
[$\frac{m^3}{\mbox{K kg}}$] at the melting point given by
\cite{iida93}. The values by Iida and Gutherie \cite{iida93} are
slightly different from the values from Miedema \cite{miedema78a}.
The uncertainty of the experimental values for the bulk surface
tension at the melting point are quoted \cite{iida93} to be $\sim
5-10\%$ at $T_m$ whereas the extrapolation towards $T=0$ resp. towards
the boiling point is done with the parameters $\frac{1}{\gamma}\frac{d
\gamma}{dT}|_{melt}$ and $\frac{1}{v}\frac{dv}{dT}|_{melt}$ which have
an estimated error of $\pm 50\%$. We give the in this way extrapolated
values of $\sigma_{bulk}(T_{boil})$ for iron in the last column of
table
\ref{table2}.  

Inspecting these numbers we find the liquid-drop parameter
$a_s=\sigma(T=0)=4\pi r_{ws}^2*\gamma|_{T=0}$ to be less by about 30\%
than the values of $a_s$ determined for the ground-state binding energies
of real clusters averaged over the shell effects in
\cite{brechignac95b} which we have of course used in our calculation
for the ground-state binding energies of the fragments.  The origin of
this experimental discrepancies is yet unknown
\cite{brechignac95b}.  Consequently, we think we should compare our theoretical
values for $\sigma_{boil}$ with values for $\sigma_{bulk}(T)$ which
are consistent at $T=0$ with the values given for $a_s$ by
\cite{brechignac95b} and than extrapolated to $T=T_{boil}$ with
formula \ref{sigmabulk}. These are listed in the last column of table
\ref{table1}.

To check this conclusion we recalculated the transition parameters for
Fe$_{1000}^+$ and Fe$_{3000}^+$ using instead of $a_s=5.1$eV,
$a_s=4.0$eV , the value of $\sigma_{bulk}(T=0)$ as estimated with the
help of formula \ref{sigmabulk} from the experimental surface tension
of the bulk at melting. Again the resulting transition parameters
listed in TABLE \ref{table2} approach nicely the corresponding
``experimental'' bulk values at $T=T_{boil}$. As our theory uses the
ground-state ($T=0$) binding energies for the clusters as input
values, the theory predicts not the total surface tension but its
temperature dependence, the so called ``entropic part of the surface
tension''. The good reproduction of the adjusted value at boiling in
TABLE \ref{table1} as well as the true one in TABLE \ref{table2} shows
the high quality of our method.

Fig. \ref{fig2} shows the same caloric curve $T_P(\varepsilon)$ as fig.\ref{fig1}
but now the number of evaporated monomers.  At $\varepsilon > 1.2$eV the
system is totally vaporised into monomers. At $\varepsilon_2\approx 0.7$eV the
character of the decay of the system changes and this can also be seen in
$N_{mon}(\varepsilon)$.
\begin{figure}
\includegraphics*[bb = 28 144 488 591, angle=-90, width=13cm,  
clip=true]{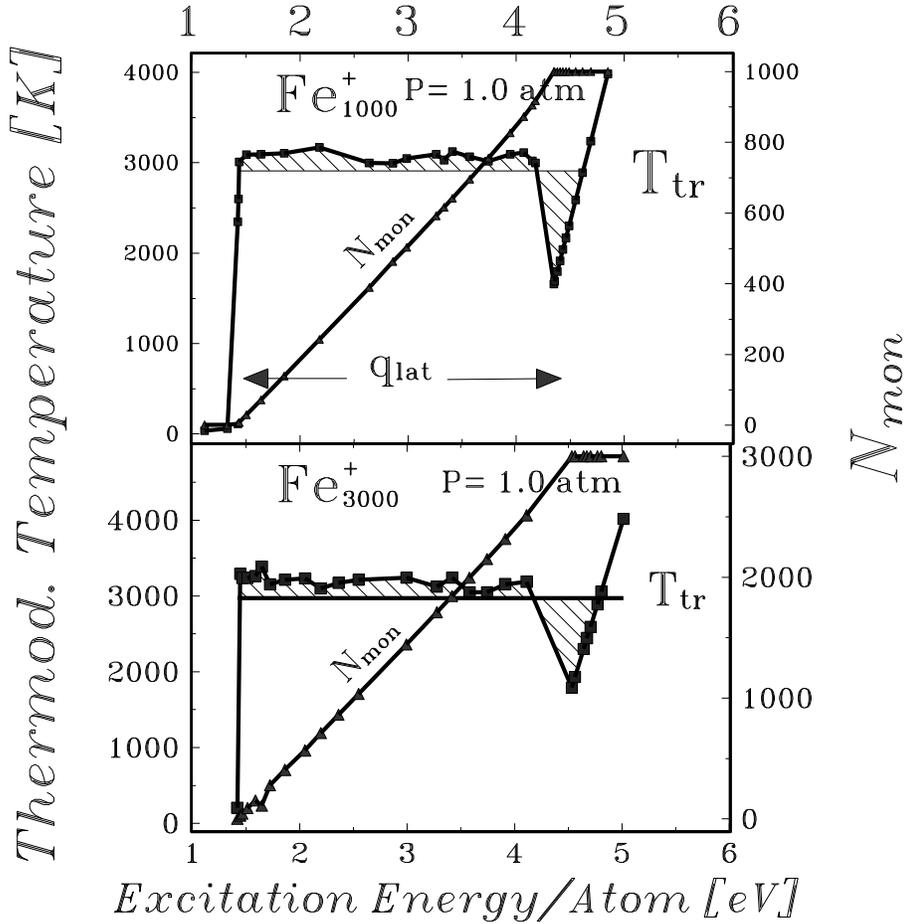}
\caption{Micro-canonical caloric curve $T_P(E/N=\varepsilon,P)$ (full square
points), number of monomers
\label{fig3}}
\end{figure}
The decay of potassium is in all details similar to that of sodium. Therefore
we don't show here the corresponding figures. The liquid--gas transition in
iron is different from that of the alkali metals: Due to the considerably
larger surface energy parameter $a_s$ in the liquid drop formula of the
ground-state binding energies of iron compared to alkali metals there is no
multi-fragmentation of iron clusters at $P=1$  atm. Iron cluster of $N\le 3000$
atoms decay by multiple monomer evaporation c.f. fig.\ref{fig3}. 
The value of $N_{eff}^{2/3}$ given in tables \ref{table1} and \ref{table2} 
for iron is taken as the average of $N_{eff}^{2/3}$ over the energy interval
$\varepsilon_1,\varepsilon_2$.

Table \ref{table1} gives a summary of all theoretical parameters for the
liquid-gas transition in clusters of $N_0=200-3000$ Na, K, and Fe atoms and
compared with their experimental bulk values. The transition-temperature
$T_{tr}$, the specific latent heat $q_{lat}$ and the entropy gain of an
evaporated atom $s_{boil}$ are well approaching the experimental bulk values. 
$\Delta s_{surf}$ is the area under the back-bend of
$\beta(\varepsilon)$. $\Delta s_{surf}*N_0$ is the total entropy loss
due to the interfaces equal to 
\begin{equation}
\sum 4\pi r_{ws}^2m_i^{2/3}N_i\gamma/T_{tr}=N_{eff}^{2/3}\sigma/T_{tr} .
\end{equation}\\~
\begin{center} 
\vspace*{-1cm}
\begin{table}[h] 
\begin{minipage}[h]{8cm}
\begin{tabular} {|c|c|c|c|c|c|} \hline 
&$N_0$&$200$&$1000$&$3000$&\vecb{bulk}\\ 
\tableline  
\hline 
\hline  
&$T_{tr} \;[K]$&$816$&$866$&$948$&\vecbm{$1156$}\\ \cline{2-6} 
&$q_{lat} \;[eV]$&$0.791$&$0.871$&$0.91$&\vecbm{$0.923$}\\ \cline{2-6} 
{\bf Na}&$s_{boil}$&$11.25$&$11.67$&$11.2$&\vecbm{$9.267$}\\ \cline{2-6} 
&$\Delta s_{surf}$&$0.55$&$0.56$&$0.45$&\\ \cline{2-6} 
&$N_{eff}^{2/3}$&$39.94$&$98.53$&$186.6$&\vecbm{$\infty$}\\
\cline{2-6} 
&$\sigma/T_{tr}$&$2.75$&$5.68$&$7.07$&\vecbm{$7.41$}\\ 
\hline 
\hline
&$T_{tr} \;[K]$&$697$&$767$&$832$&\vecbm{$1033$}\\ \cline{2-6}
&$q_{lat} \;[eV]$&$0.62$&$0.7$&$0.73$&\vecbm{$0.80$}\\ \cline{2-6}
{\bf K}&$s_{boil}$&$10.35$&$10.59$&$10.15$&\vecbm{$8.99$}\\ \cline{2-6}
&$\Delta s_{surf}$&$0.65$&$0.65$&$0.38$&\\ \cline{2-6}
&$N_{eff}^{2/3}$&$32.52$&$92.01$&$187$&\vecbm{$\infty$}\\ \cline{2-6}
&$\sigma/T_{tr}$&$3.99$&$7.06$&$6.06$&\vecbm{$7.31$}\\
\hline
\hline 
&$T_{tr} \;[K]$&$2600$&$2910$&$2971$&\vecbm{$3158$}\\ \cline{2-6} 
&$q_{lat} \;[eV]$&$2.77$&$3.18$&$3.34$&\vecbm{$3.55$}\\ \cline{2-6} 
{\bf Fe}&$s_{boil}$&$12.38$&$12.68$&$13.1$&\vecbm{$13.04$}\\ \cline{2-6} 
&$\Delta s_{surf}$&$0.75$&$0.58$&$0.77$&\\ \cline{2-6} 
&$N_{eff}^{2/3}$&$22.29$&$65.40$&$142.12$&\vecbm{$\infty$}\\
\cline{2-6} 
&$\sigma/T_{tr}$&$6.73$&$8.87$&$16.25$&\vecbm{$17.49$}\\ 
\hline 
\end{tabular}  
\end{minipage}\\~\\
\caption{Parameters of the liquid--gas transition at constant pressure 
of $1$atm. in a micro-canonical system of $N_0$ interacting atoms and
in the bulk. $s_{boil}=q_{lat}/T_{tr}$, it is interesting that the
value of $s_{boil}$ for all three systems and at all sizes is near to
$s_{boil}=10$ as proposed by the empirical Trouton's rule
\protect\cite{reif65}, $N_{eff}^{2/3}=\sum{m_i^{2/3} N_i}$ and
$\sigma/T_{tr}=N_0\Delta s_{surf}/N_{eff}^{2/3}$.  The bulk values
$\sigma/T_{tr}$ are adjusted according to formula \ref{sigmabulk} to
the input values of $a_s$ taken for the $T=0$ liquid-drop surface
parameters from ref.\protect\cite{brechignac95b} which we used in the
present calculation for the ground-state binding energies of the
fragments.
\label{table1} } 
\end{table} 
\end{center} 
\newpage
\begin{center} 
\begin{table}[h] 
\begin{minipage}[h]{6cm}
\begin{tabular} {|c|c|c|c|c|} \hline 
&$N_0$&$1000$&$3000$&\vecb{bulk}\\ 
\tableline  
\hline 
&$T_{tr} \;[K]$&$2994$&$3044$&\vecbm{$3158$}\\ \cline{2-5} 
&$q_{lat} \;[eV]$&$3.13$&$3.27$&\vecbm{$3.55$}\\ \cline{2-5} 
{\bf Fe}&$s_{boil}$&$12.13$&$12.47$&\vecbm{$13.04$}\\ \cline{2-5} 
&$\Delta s_{surf}$&$0.48$&$0.39$&\\ \cline{2-5} 
&$N_{eff}^{2/3}$&$65.74$&$136.78$&\vecbm{$\infty$}\\
\cline{2-5} 
&$\sigma/T_{tr}$&$7.30$&$8.62$&\vecbm{$8.86$}~\\ 
\hline 
\end{tabular}  
\end{minipage}\\~\\~\\
\caption{Same as table \protect\ref{table1}, but using for the ground-state 
liquid-drop parameter $a_s$ the value
compiled by extrapolating the experimental surface tension of bulk
iron at melting down to $T=0$ (c.f. the discussion in the text) or to
$T=T_{boil}$. \label{table2} } 
\end{table} 
\end{center}
Of course, the transition temperature $T_{tr}$ and the latent
heat $q_{lat}$ of small clusters are smaller than the bulk values because
the average coordination number of an atom at the surface of a small
cluster is smaller than at a planar surface of the bulk.  It is
somewhat surprising that the surface tension is rising from $N_0=200$
to $N_0=3000$. One might think it should fall as $\sigma$ is expected
to be proportional to the number of broken bonds per surface
area. However, this is a purely static argument and does not take the
entropic (fluctuational) part of $\sigma$ into account. 
\section{The relation to the method of the Gibbs-ensemble}  
The alternative method to simulate the liquid--gas transition within the
canonical ensemble is by the method of the Gibbs-ensemble
\cite{panagiotopoulos87}. As configurations with two coexistent phases
separated by an interface become asymptotically suppressed, see above, in the
Gibbs--ensemble method the system is sampled in two separated containers which
can communicate via particle and energy exchange. In one container the system
is close to the bulk liquid density and specific energy and in the other it is
in the corresponding gas phase  without establishing an inner, and possibly
fluctuating interface. Then the transition temperature and latent heat can be
determined.

The Gibbs-ensemble method is used to investigate the critical behaviour of
systems interacting via short range forces like Lennard-Jones potentials,
c.f. the recent article by Bruce \cite{bruce97}.

The surface tension cannot be determined by the Gibbs-ensemble method.
The dramatic fluctuations of the physical interfaces between both phases at
$T=T_{tr}$ which can be seen e.g. in boiling water are a clear warning of
the much more complicated and strongly fluctuating microscopic topology of the
interfaces. They are considerably more complicated than the geometry assumed in
the Gibbs-ensemble method.  The surface tension is proportional to the entropy
of the interface-fluctuations and consequently it depends essentially on these
fluctuations.

\section{Alternative microscopic methods to calculate the surface tension}
Microscopic canonical calculations of the surface tension have been
done for Lennard--Jones systems.  A detailed survey is given by
Salomons and Mareschal \cite{salomons91}.  They discuss the following
alternatives to determine the surface tension:
\begin{enumerate}
\item from the virial expansion of the free-energy as function of volume
conserving distortions of the interface:
\begin{equation}
\gamma(T)=\frac{1}{2A}\left<\sum_{i<j}\left(1-\frac{3z_{ij}^2}{r_{ij}^2}r_{ij}\phi'
(r_{ij})\right)\right>,
\end{equation}
where $r_{ij},z_{ij}$ are the interatomic coordinates of the atoms,
$A$ is the surface area and $\phi'(r_{ij})$ is the derivative of the
interatomic {\em two-body} potential,
\item from histograms of the free-energy as function of potential energy 
changes due to volume conserving deformations.
\end{enumerate}
The authors of \cite{salomons91} also determine the ``entropic'' part of the
surface tension $\partial \gamma/\partial T$ either directly from a compilation
of $\gamma(T)$ at various temperatures or via the fluctuation equation
\begin{equation}
\frac{d\gamma}{dT}=\frac{1}{T^2}<(\hat{U}-U)\gamma>,
\end{equation}
where $U$ is the total potential energy and $\hat{U}$ its mean
value.

Evidently, the virial expansion works only for systems with two-body forces.
So we cannot use it for metallic systems. The second method has not been tried
either.
\section{Criticism and necessary improvements of the computational method}
Several simplifications were made:
\begin{itemize}
\item We use classical not quantum statistics. --- We do not believe that the
problems we discussed in the present paper depend essentially on that
difference. 
\item The fragments were assumed to be spherical. --- This simplification means
that surface degrees of freedom of the fragments are treated like bulk
excitations. The size of the fragments was assumed to be independent
of the excitation energy at $\varepsilon\le\varepsilon_{max}$.
Moreover, the static deformation of dimers and trimers is quite
important for decay thresholds. An improvement of these approximations
was not possible in the present calculation.
\item Only internal excitations of the fragments of $\varepsilon \le
\varepsilon_{max}$ were allowed. --- This approximation may be justified for
the decay of hot finite clusters \cite{gross141,gross153} where we
argued that only a transient equilibrium of the decaying cluster over
a time $\tau_{equ}$ is achieved and states which live considerably
shorter should not contribute as independent decay channels. This
argument does not work in the present scenario.  However, we believe
that highly unbound states above the boiling excitation
$\varepsilon_{boil}$ in small fragments are anyway better described by
assuming an independent motion of the decay products, just as we do it
here. Doubling the actual value of $\varepsilon_{max}$ had in most
cases only little influence on the transition parameters obtained.
\item The calculation of the total intra-phase surface {\em area} is
problematic.  Consequently, the determination of the surface tension per
surface area $\gamma$ suffers under a great uncertainty, whereas the
determination of the surface {\em entropy} $\Delta s_{surf}$ is easy. The
situation is complementary to the Gibbs ensemble method
\cite{panagiotopoulos87}. However, the surface entropy is crucial for
the occurrence and classification of a phase transition, not its
geometrical interpretation.
\item The input values of the liquid drop parameters of the ground-state 
binding energies of the fragments given by \cite{brechignac95b} suffer
from the uncertainty of the temperature of the clusters where the 
experimental data were taken. The bulk measurements of the surface
tension suffers from many effects c.f.\cite{allen85,iida93}:
\begin{itemize}
\item The surface tension of alkali clusters might be reduced by oxygen
contamination.
\item The volume expansion coefficient $\frac{1}{v}dv/dT$ is only known
within $\pm50\%$.
\item $\gamma(T=0)$ refers to the solid phase which is most likely anisotropic
and cannot easily be extrapolated from the melting point.
\end{itemize}
\end{itemize}  
\section{Conclusion}
The liquid--gas transition in metals at normal pressure of 1 atm. is
experimentally well explored. Therefore, it is a good test case for
our new ideas and computational methods of micro-canonical
thermodynamics. By allowing a system of $N=200-3000$ atoms to condense
or fragment into an arbitrary number of spherical fragment clusters
which are internally excited to energies of $\varepsilon \le
\varepsilon_{boil}$ per atom and into an arbitrary number of free
atoms under a fixed given external pressure of 1 atmosphere we
directly calculated the micro-canonical caloric curve
$\beta(\varepsilon)=\partial s/\partial
\varepsilon=<\partial/\partial \varepsilon>$ (c.f. the detailed formula 
eq.(\ref{betap}))
 by micro-canonical Monte Carlo methods ({\em MMMC})\cite{gross153}. The
 volume $W(E)$ of the total accessible N-body phase space
 can then be determined from $\beta_P(\varepsilon)$  by
 $s(\varepsilon)=\int^{\varepsilon}\beta(\varepsilon')d\varepsilon'$
 and $W(E)=\mbox{exp}[Ns( E/N)]$. The anomaly of the micro-canonical
 caloric curve $T_P(\varepsilon)$ signals the liquid--gas
 phase-transition in the finite system. The characteristic parameters
 $T_{tr}$, $q_{lat}$, and the surface tension $\sigma_{surf}$ approach
 already for $\sim 1000 - 3000$ atoms the experimentally known values
 of the bulk liquid--gas transition.

This result is remarkable for several reasons:
\begin{enumerate}
\item It proves that a phase transition of first order in a realistic
continuous system can very well be seen and classified in small
mesoscopic clusters {\em without invoking the thermodynamic limit.} In
fact, with the non-vanishing back-bending of $T_P(E/N,V(E,N,P))$ the transition
is easier recognisable than in the thermodynamic limit. For the
10-states Potts model we know that the micro-canonical parameters
depend less on the size of the system than the canonical ones
\cite{gross150}. Finite size scaling for the micro-canonical
transition parameters are not known in general. However, here we want
to stress that there is no reason at all not to discuss phase transitions
in small systems.
\item Even though the values determined for the specific surface tension
$\sigma=4\pi r_{ws}^2\gamma$ are somewhat uncertain because of the difficulty
to fix and determine the surface {\em area}, the surface {\em entropy} $N\Delta
s_{surf}$ which is the relevant quantity that determines the nature of the
transition can very well be calculated.
\item Even for a realistic metallic system with its long-range many-body 
interactions the {\em M}icrocanoncal {\em M}etropolis {\em M}onte {\em
C}arlo simulation method ({\em MMMC}) is able to describe the liquid--gas phase
transitions quite well.  This is possible because we do not use molecular
dynamics with a two-body Hamiltonian but use the experimental ground-state
binding energies of the fragment clusters which of course take care of the
metallic bonding of their constituents.
\item The intra-phase surface entropy can be microscopically calculated. 
(Essentially we calculated the temperature dependent part of it. The value
at $T=0$ is an input parameter of our calculation.) The 
surface tension per surface area can be determined if the intra-phase
{\em area} is known. For the surface entropy the fluctuation and
fragmentation of the surfaces are essential. At the liquid--gas
transition of sodium and potassium clusters of sizes as considered
here at normal pressure the intra-phase fluctuations are mainly due to
strong inhomogeneities and clusterization and not due to homogeneous
stretching. This is consistent with recent experimental evidence
\cite{hensel95,ross96}.
\item Micro-canonical thermodynamics gives an important insight into the details
of the transition which are hidden in the canonical approach. The transition
in sodium is a good example: Figs.\ref{fig1} and \ref{fig1a} show how the
transition is a combination of actually two transitions, one from evaporation
of very light fragments out of a large residue into a multi-fragmentation of the
system into many larger pieces, and second the complete decay into a gas of
monomers.
\item The success of the {\em M}icrocanoncal {\em M}etropolis {\em M}onte {\em
C}arlo sampling method to reproduce the known infinite matter values
of the liquid--gas transition is also a promising and necessary test
of {\em MMMC} to describe nuclear fragmentation correctly
\cite{gross95,gross153} and to be able to get insight into the
(eventually critical) behaviour of the nuclear matter liquid--gas
transition. This is important as an experimental test of the
predictions of the model for nuclear matter is not available.
\item The division of the macroscopic time-independent observables of 
the N-body system into
\begin{enumerate}
\item
observables which can be determined at each phase-space {\em point},
in each individual realisation of the micro-canonical ensemble: The
globally conserved quantities like the energy, the number of
particles, the charge, the momentum, the angular momentum, and
\item
into the thermodynamic observables which refer to the size of the ensemble,
the {\em volume} $e^S$ of the energy shell of the N-body phase space, and which
cannot be determined at a single phase-space point, in a single event, like the
entropy, the temperature, the pressure, the chemical potential
\end{enumerate}
is very essential. Also the concept of a pure phase and of a phase-transition
belongs to the second group.
\item Last not least, there is a very fundamental difference between
micro-canonical and canonical ensembles. This cannot be emphasised
enough: As said above, if one wants to study a configuration where
two-phases coexist simultaneously and are separated by an interface one
has to control energy (if pressure is fixed). Thus the {\em the micro
ensemble \{E,P,N\} is the adequate ensemble at phase transitions of
first order in the bulk, not the canonical}, at least if the energy
resources are not unlimited as they usually are.
\end{enumerate}
A preliminary version of this paper was already published in \cite{gross158}.
\section{Acknowledgement}
D.H.E.G. is grate-full to S.Gro\ss mann for emphasising the intimate
relation of the anomaly of the micro-canonical curve $T(E)$ to phase
transitions of first order more than 15 years ago.  We are also
grateful to M.Gross for putting our attention to the definition of the
temperature as an ensemble average which is the source of many
conceptional difficulties for temperature-measurements in small closed
systems. Thanks to A.Ecker for helping us in the use of MAPLE. We
thank the SFB337 of the Deutsche Forschungsgemeinschaft (DFG) for
granting a postdoc position for M.Madjet.

\begin{thebibliography}{10}

\bibitem{boltzmann}
Ludwig Boltzmann.
\newblock {\em Vorlesung \"uber Gastheorie}.
\newblock Vol~1. Akademische Druck-u. Verlagsanstalt, Graz, 1981.

\bibitem{gross154}
D.H.E. Gross, M.E. Madjet, and O.~Schapiro.
\newblock Fragmentation phase transition in atomic clusters I ---
  microcanonical thermodynamics.
\newblock {\em Z.Phys.D}, 39:75--83;http://xxx.lanl.gov/cond--mat/9608103,
  1997.

\bibitem{gross151}
M.E. Madjet, D.H.E. Gross, P.A. Hervieux, and O.~Schapiro.
\newblock Fragmentation phase transition in atomic clusters II --- symmetry of
  coulombic fission.
\newblock {\em Z.Physik D}, 39:309--316;http://xxx.lanl.gov/cond--mat/9610118,
  1997.

\bibitem{gross152}
O.~Schapiro, P.J. Kuntz, K.~M\"ohring, P.A. Hervieux, D.H.E. Gross, and M.E.
  Madjet.
\newblock Fragmentation phase transition in atomic clusters III --- coulomb
  explosion of metal clusters.
\newblock {\em Z.Physik D in print}, http://xxx.lanl.gov/cond-mat/9702183,
  1997.

\bibitem{gross72}
D.H.E. Gross, X.Z. Zhang, and S.Y. Xu.
\newblock Decay of very hot nuclei.
\newblock {\em Phys. Rev. Lett}, 56:1544, 1986.

\bibitem{gross75}
X.Z. Zhang, D.H.E. Gross, S.Y. Xu, and Y.M. Zheng.
\newblock Decay of very hot nuclei,II, microcanonical Metropolis sampling of
  multifragmentation.
\newblock {\em Nucl. Phys.}, A~461:668--690, 1987.

\bibitem{gross95}
D.H.E. Gross.
\newblock Statistical decay of very hot nuclei, the production of large
  clusters.
\newblock {\em Rep.Progr.Phys.}, 53:605--658, 1990.

\bibitem{challa88}
M.S.S. Challa and J.H. Hetherington.
\newblock Gaussian ensemble: An alternate Monte Carlo scheme.
\newblock {\em Phys.Rev. A}, 38:6324, 1988.

\bibitem{gross124}
D.H.E. Gross and R.~Heck.
\newblock What is wrong with the Bethe formula ? - Measurable differences
  between the grandcanonical and microcanonical ensemble.
\newblock {\em Phys. Lett.B}, 318:405--409, 1993.

\bibitem{bixton89}
M.~Bixton and J.~Jortner.
\newblock Energetic and thermodynamic size effects in molecular clusters.
\newblock {\em J.Chem.Phys.}, 91:1631--1642, 1989.

\bibitem{grimsditch96}
M.~Grimsditch and V.G. Karpov.
\newblock Fluctuations during melting.
\newblock {\em J. Phys.:Condens.Matter}, 8:L439--L444, 1996.

\bibitem{gross150}
D.H.E. Gross, A.~Ecker, and X.Z. Zhang.
\newblock Microcanonical thermodynamics of first order phase transitions
  studied in the Potts model.
\newblock {\em Ann. Physik}, 5:446--452, 1996.
\bibitem{thirring70}

W.~Thirring.
\newblock Systems with negative specific heat.
\newblock {\em Z. f. Phys.}, 235:339--352, 1970.

\bibitem{vanhove49}
L.~van Hove.
\newblock Quelques propri$\acute{e}$t$\acute{e}$s g$\acute{e}$n$\acute{e}$rales
  de l'int$\acute{e}$grale de configuration d\'\,un syst$\grave{e}$me de
  particules avec interaction.
\newblock {\em Physica}, 15:951, 1949.

\bibitem{bohrmott}
Aa. Bohr and B.R. Mottelson.
\newblock {\em Nuclear Structure}.
\newblock W.A. Benjamin, New York, Amsterdam, 1969.

\bibitem{hueller94}
A.~H\"uller.
\newblock Finite size scaling at first order phase transitions ?
\newblock {\em Z.Phys.B}, 95:63--66, 1994.

\bibitem{promberger96}
M.~Promberger.
\newblock On a trivial aspect of canonical specific heat scaling.
\newblock {\em preprint, Erlangen}, 1996.

\bibitem{ruelle69}
D.~Ruelle.
\newblock {\em Statistical Mechanics, Rigorous Results}.
\newblock W.A.Benjamin, New York, Amsterdam, 1969.

\bibitem{panagiotopoulos87}
A.~Z. Panagiotopoulos.
\newblock {\em Mol.Phys.}, 61:813, 1987.

\bibitem{evans92}
R.~Evans.
\newblock Density functionals in the theory of nonuniform fluids.
\newblock In D.~Henderson, editor, {\em Fundamentals of Inhomogeneous Fluids},
  chapter~3, pages 85--200. M. Dekker, Inc, New York, 1992.

\bibitem{gross153}
D.H.E. Gross.
\newblock Microcanonical thermodynamics and statistical fragmentation of
  dissipative systems --- the topological structure of the n-body phase space.
\newblock {\em Physics Reports}, 279:119--202, 1997.

\bibitem{andersen80}
H.C. Andersen.
\newblock {\em J.Chem.Phys.}, 72:2284, 1980.

\bibitem{brown84}
D.~Brown and J.H.R. Clarke.
\newblock A comparison of constant energy, constant temperature and constant
  pressure ensembles in molecular dynamics simulations of atomic liquids.
\newblock {\em Mol.Phys.}, 51:1243--1252, 1984.

\bibitem{procacci94}
P.~Procacci and B.J. Berne.
\newblock Multiple time scale methods for constant pressure molecular dynamics
  simulations of molecular systems.
\newblock {\em Mol.Phys.}, 83:255--272, 1994.

\bibitem{allen85}
B.C. Allen.
\newblock Surface tension.
\newblock In R.W. Ohse, editor, {\em Handbook of Thermodynamic and Transport
  Properties of Alkali Metals}, chapter 6.8, pages 691--700. Blackwell
  Scientific, Oxford, 1985.

\bibitem{gross141}
D.H.E. Gross and P.A. Hervieux.
\newblock Statistical fragmentation of hot atomic metal clusters.
\newblock {\em Z. Phys. D}, 35:27--42, 1995.

\bibitem{hensel95}
F.~Hensel.
\newblock The liquid..vapour phase transition in fluid mercury.
\newblock {\em Advances in Physics}, 44:3--19, 1995.

\bibitem{ross96}
M.~Ross and F.~Hensel.
\newblock A modified van der waals model for the coexistence of expanded
  metals.
\newblock {\em J.Phys.: Condens.Matter}, 8:1909--1919, 1996.

\bibitem{iida93}
T.~Iida and R.I. Gutherie.
\newblock {\em The Physical Properties of Liquid Metals}, chapter~5.
\newblock Clarendon Press, Oxford, 1993.

\bibitem{miedema78a}
A.~Miedema and R.~Boom.
\newblock Surface energies and electron density of pure liquid metals.
\newblock {\em Z.Metallkde}, 69:183, 1978.

\bibitem{brechignac95b}
C.~Br\'echignac, Ph. Cahuzac, F.~Carlier, M.~de~Frutos, J.~Leygnier, J.Ph Roux,
  and A.~Sarfati.
\newblock Dissosiation and binding energies of metal clusters.
\newblock {\em Comments At.Mol.Phys.}, 31:361--393, 1995.

\bibitem{reif65}
F.~Reif.
\newblock {\em Fundamentals of statistical and thermal physics}.
\newblock McGraw-Hill, New York, 1965.

\bibitem{bruce97}
A.D. Bruce.
\newblock Finite-size critical behavior in the Gibbs ensemble.
\newblock {\em Phys.Rev.E}, 55:2315--2320, 1997.

\bibitem{salomons91}
E.~Salomons and M.~Mareschal.
\newblock Surface tension, adsorption and surface entropy of liquid--vapour by
  atomistic simulation.
\newblock {\em J.Phys.: Condens.Matter}, 3:3645--3661, 1991.

\bibitem{gross158}
D.H.E. Gross and M.E. Madjet.
\newblock Microcanonical vs. canonical thermodynamics.
\newblock {\em http://xxx.lanl.gov/cond-mat/9611192}.

\end{thebibliography}

\end{document}